\documentclass[12pt]{article}

\newcommand{\sect}[1]{\noindent{\bf {#1}.} }

\newcommand{\ff}{{\rm f} }
\newcommand{\vn}{{\bf n}}
\newcommand{\vm}{{\bf m}}
\newcommand{\vl}{{\bf l}}

\newcommand{\vv}{V}
\newcommand{\vvv}{v}
\newcommand{\cG}{{\cal G}}
\newcommand{\starx}{\star}

\newcommand{\A}{{\sf A}}
\newcommand{\cD}{{\cal D}}

\newcommand{\ee}[1]{{\rm e}^{#1}}

\newcommand{\eq}{\begin{equation}}
\newcommand{\eqend}{\end{equation}}
\newcommand{\eqa}{\begin{eqnarray}}
\newcommand{\nonueqa}{\begin{eqnarray*}}
\newcommand{\eqaend}{\end{eqnarray}}
\newcommand{\nonueqaend}{\end{eqnarray*}}
\newcommand{\nonu}{\nonumber \\ \nopagebreak}
\newcommand{\bma}[1]{\begin{array}{#1}}
\newcommand{\ema}{\end{array}}
\newcommand{\bc}{\begin{center}}
\newcommand{\ec}{\end{center}}

\newcommand{\Ref}[1]{(\ref{#1})}

\newcommand{\R}{\real}

\def\appendix#1{\addtocounter{section}{1}\setcounter{equation}{0}
\renewcommand{\thesection}{\Alph{section}}
\section*{Appendix \thesection\protect\indent \parbox[t]{11.715cm} {#1}}
\addcontentsline{toc}{section}{Appendix \thesection\ \ \ #1} }

\newcommand{\zed}{{\bb Z}} 
\newcommand{\real}{{\bb R}} 
\newcommand{\reals}{{\bbs R}} 

\newif\ifold             \oldtrue

\font\mybb=msbm10 at 12pt
\def\bb#1{\hbox{\mybb#1}}

\font\mybbs=msbm10 at 9pt
\def\bbs#1{\hbox{\mybbs#1}}

\def\e{{\,\rm e}\,}

\def\be{\begin{equation}}
\def\ee{\end{equation}}
\def\bea{\begin{eqnarray}}
\def\eea{\end{eqnarray}}
\def\bd{\begin{displaymath}}
\def\ed{\end{displaymath}}

\newcommand{\beq}{\begin{eqnarray}}
\newcommand{\eeq}{\end{eqnarray}}

\begin{document}

\begin{flushright}
\baselineskip=12pt
HWM--02--3\\
EMPG--02--03\\
hep--th/02020039\\
\hfill{ }\\
February 2002
\end{flushright}

\begin{center}

\baselineskip=20pt

{\Large\bf Duality in scalar field theory\\ on noncommutative phase spaces}

\baselineskip=12pt

\vspace{0.5 cm}

{\large Edwin Langmann$^a$ and Richard J.\ Szabo$^b$}\\

\vspace{0.4 cm}

{\it $^a$ Department of Physics -- Mathematical Physics, Royal Institute of
Technology\\Stockholm Center for Physics, Astronomy and Biotechnology\\ S-10691
Stockholm, Sweden\\{\tt langmann@theophys.kth.se}}\\[3mm]

{\it $^b$ Department of Mathematics, Heriot-Watt University\\ Riccarton,
Edinburgh EH14 4AS, Scotland\\{\tt R.J.Szabo@ma.hw.ac.uk}}

\end{center}

\begin{abstract}

\baselineskip=12pt

\noindent
We describe a novel duality symmetry of $\Phi_{2n}^4$-theory defined on
noncommutative Euclidean space and with noncommuting momentum coordinates. 
This duality acts on the fields by Fourier transformation and scaling. 
It is an extension, to interactions defined with a star-product, of
that which arises in quantum field theories of non-interacting
scalar particles coupled to a constant background electromagnetic field. The
dual models are in general of the same original form but with
transformed coupling parameters, while in certain special cases  
all parameters are essentially unchanged. Using a particular regularization
we show, to all orders of perturbation theory, that this duality also
persists at the quantum level. We also point out various other
properties of this class of noncommutative field theories.

\end{abstract}

\baselineskip=14pt

\sect{1. Introduction} Conventional particle and condensed matter models are
often based upon quantum field theories with local
interactions. Recently, however, 
a novel class of non-local field theories has come to the center of
attention. These models can be obtained from the standard local ones
by defining the contact interactions, which in the standard cases are
given by local products of the fields, using the non-local
Groenewold-Moyal star-products. The remarkable properties of these
models have been under extensive investigation in a variety of
different contexts (see~\cite{NCrev} for reviews and fairly exhaustive
lists of references). One reason for this interest is that these
models have a natural interpretation in terms of fields living on a
noncommutative spacetime $\R^{2n}$ in which the coordinates
$x=(x^1,x^2,\ldots,x^{2n})=(x^\mu)$ obey the (star-) commutation
relations
\eq x^\mu\star x^\nu - x^\nu \star x^\mu = -2i\,\theta^{\mu\nu} \ , \eqend
where $\theta=(\theta^{\mu\nu})$ is a constant $2n\times2n$ skew-symmetric real
matrix. These models also naturally emerge as low-energy effective field
theories in string theory with constant background electromagnetic
fields~\cite{SW}.

A simple example of such a class of models is provided by noncommutative
$\Phi^4_{2n}$-theories, i.e.\ bosons on $\R^{2n}$ with a quartic interaction in
which the local products of the fields are replaced by star-products. The
generalization of field theories to ones defined on noncommutative spacetimes
naturally suggests the generalization which allows for noncommutative momentum
spaces as well. This modification may also be motivated by the
observation~\cite{SW,Morita} that such extensions lead, in the case of
noncommutative Yang-Mills theory, to actions which are explicitly invariant
under Morita duality transformations and they allow one to interpolate between
commutative and noncommutative descriptions of the same theory. For scalar
field theories it is achieved
by coupling the (complex) bosons to an external, constant magnetic field
defined by another skew-symmetric $2n\times2n$ matrix $B=(B_{\mu\nu})$. Similar
models have emerged as effective descriptions of some planar condensed matter
systems in strong magnetic fields~\cite{Hall}, such as quantum Hall 
models. A similar relativistic scalar field theory has been studied
in~\cite{Habara}, and its one-particle sector, i.e.\ the noncommutative
Landau problem, in~\cite{NCLandau}. A further natural generalization
allows for an arbitrary flat metric $G=(G_{\mu\nu})$ on $\real^{2n}$
determined by a constant, positive-definite symmetric matrix. In this
paper we will show that this class of quantum field theories has a
remarkable duality property under Fourier transformation, and we make
various other related observations. We will always assume that the
matrices $B$ and $\theta$ are invertible.

Before stating this duality property of the noncommutative scalar field theory,
we recall a well-known toy version of it which occurs in the quantum mechanical
harmonic oscillator. Consider the differential operator
\eq
H=-\frac{\partial^2}{\partial y^2} + \omega^2\,y^2\equiv H(y;\omega) \ ,
\eqend
where $\omega>0$ and $y$ is a coordinate on the real line. Clearly, under
Fourier transformation in $y$ it becomes
\eq
\hat H(p;\omega) = -\omega^2\,\frac{\partial^2}{\partial p^2} + p^2
= -\frac{\partial^2}{\partial (p/\omega)^2} + \omega^2\,(p/\omega)^2 \ ,
\eqend
where $p$ is the corresponding momentum space coordinate, or equivalently, 
\eq
\label{1}
\hat H(p;\omega) = \omega^2\,H\left(p\,;\,\omega^{-1}\right) =
H\left(\omega^{-1}\,p\,;\,\omega\right) \ .
\eqend
This fact can be used to explain the special property of the harmonic
oscillator ground state wavefunction that its Fourier transform equals itself,
up to a rescaling. This is also 
the simplest example of a strong-weak coupling duality in a quantum theory.

In a similar manner, we will find that the quantum field theories of
charged bosons on noncommutative spacetime and with noncommutative
momentum space, parametrized by the matrices $G$, $B$ and $\theta$ as
described above, and a coupling constant 
$g$, preserve their form under Fourier
transformation.\footnote{\baselineskip=12pt To avoid confusion we
point out that, after introducing the magnetic field and the
star-product in the action of this model, all field theory
computations are done in terms of standard (commutative) position and
Fourier variables, and what we mean here is standard Fourier
transformation. The noncommutative spaces alluded to above merely
provide a useful interpretation of these models.} More specifically,
Fourier transformation and a rescaling of the fields amounts to
changing the parameters of the field theory
as\footnote{\baselineskip=12pt
Our conventions for $B$, $\theta$ and $g$ are defined in Section~2.}
\eqa
\label{11}
G&\longmapsto& G \ , \nonu
B&\longmapsto& B \ , \nonu
\theta&\longmapsto&-B^{-1}\,\theta^{-1}\,B^{-1} \ , \nonu
g&\longmapsto&\Bigl|\det(B\theta)\Bigr|^{-1/2}\,g \ .
\eqaend
This property remains true even if one adds a mass term for the scalar fields
and allows for both types of quartic star-interaction terms which are possible
for charged bosons~\cite{ABK00}. Then the mass and the relative weights of
these interactions are invariant under this transformation. We interpret
Fourier transformation followed by this specific rescaling of the fields (for
details see Proposition~1 below) as a `duality' transformation of the field
theory. It maps the model to one of the same form but with some transformed
`coupling parameters' (the parameters characterizing the interaction). Note
that the free part of the action, parametrized by the metric, the magnetic
field, and the mass, are unchanged by this transformation, and it is consistent
to restrict to the case of a trivial metric $G=I$ (the identity matrix). This
duality can be regarded as a generalization of the second equality in the
harmonic oscillator identity~\Ref{1}. There is also an analog of the first
equality which we will discuss in Section~4 (see~\Ref{FT}). It is interesting
to note that there are two special points in parameter space at $\theta
B=\pm\,I$ where the coupling parameters are invariant up to a change in sign of
$\theta$. For each coupling constant $g$, 
the duality maps the field theory with $\theta(B)=+B^{-1}$ to that with
$\theta(B)=-B^{-1}$ and thus identifies the two special points. As we
will discuss, these `self-dual' models have remarkable special
properties. The significance of the points $\theta B=\pm\,I$ has been
noted in different contexts in~\cite{SW,Morita,NCLandau}.

We will prove this duality first at the level of the classical action by
straightforward calculation. For the free part, this property is in fact proven
by a simple extension of the argument above demonstrating the duality of the
harmonic oscillator. The remarkable feature of the duality is that the
interaction given by the star-product also possesses this property. We will
then argue that this duality carries over to the full quantum field theory. All
Green's functions of this model have the property that Fourier transformation
is the same as performing the replacements of parameters in \Ref{11}, replacing
the spacetime coordinates $x$ by rescaled Fourier variables $B^{-1}\,k$, and
multiplying by $|\det(B)|^{N/2}$, where $N$ is the number of external legs of
the given Green's function. In particular, the `self-dual' models define
correlation functions which are invariant under Fourier transformation and
rescaling. The crucial point in this extension to the full quantum field theory
is the existence of a regularization which is preserved by the duality
transformation.

In the next section we give precise definitions of the field theories that we
consider in this paper and the proof of the duality at the level of the
classical action. In Section~3 we extend the proof to the full quantum field
theory. In Section~4 we summarize our results and point out some further
interesting properties of these models. Some technical details of the proofs
are deferred to two appendices at the end of the paper.

\bigskip

\sect{2. The classical action} We consider the field theory of a charged,
massive scalar field $\Phi$ in Euclidean even-dimensional spacetime $\R^{2n}$
defined by the classical action $S= S_0+g^2\,S_{\rm int}$. The free part of
the action is given by
\eqa S_0=\int d^{2n}x~\sqrt{\det (G) } \,\left[
\left(G^{-1}\right)^{\mu\nu}\,(P_\mu \Phi)^\dag(x)(P_\nu\Phi)(x) + m^2\,
\Phi^\dag(x)\Phi(x) \right] \ ,
\label{S0}\eqaend
where $m$ is the mass parameter, and
\eqa
\label{P}
(P_\mu\Phi)(x)\equiv\left(-i\,\partial_\mu-B_{\mu\nu}\,x^\nu\right)\Phi(x)
\label{Pmudef}\eqaend
with $\partial_\mu=\partial/\partial x^\mu$. The interaction part is
\eq
S_{\rm int} = \int d^{2n}x~\sqrt{\det (G) }\,\left[
\alpha\,(\Phi^\dag\starx\Phi\starx \Phi^\dag\starx\Phi)(x)+\beta\,
(\Phi^\dag\starx\Phi^\dag\starx\Phi\starx\Phi)(x)\right] \ ,
\label{Sint}\eqend
where
\eq
(f_1\starx f_2)(x) =
f_1(x)\,\exp\left(-i\,\overleftarrow{\partial}_\mu
\,\theta^{\mu\nu}\,\overrightarrow{\partial}_\nu \right)\,f_2(x)
\eqend
is the Groenewold-Moyal star-product. Note that we differ by factors
of 2 from the usual conventions for $B$ and $\theta$, in order to
simplify some of the formulas which follow. The action $S_0$ describes
scalar bosons of mass $m$ coupled to a constant external
electromagnetic field $F_{\mu\nu}=2B_{\mu\nu}$. Since
$[P_\mu,P_\nu]=-2i\,B_{\mu\nu}$, one can also interpret $B$ as a
parameter which produces noncommuting momentum space coordinates. In
the action $S_{\rm int}$ we have included the two inequivalent,
noncommutative quartic interactions of a complex scalar
field~\cite{ABK00} which we weight by the real
parameters $\alpha$ and $\beta$.

We will now give a precise formulation of the duality in the classical field
theory.

\bigskip
\noindent {\bf Proposition~1 (Classical duality):} {\it The
action\footnote{\baselineskip=12pt We assume here that $\Phi$ is a Schwartz test function on $\reals^{2n}$ for simplicity.
In the notation we explicitly indicate only
the parameters $B$, $g$ and $\theta$ which are affected by or are involved in
the duality transformations and suppress the dependence on $G$, $\alpha$,
$\beta$ and $m$ which remain unchanged. The notation $S[\Phi;\ldots]$ is
short-hand for $S[\Phi,\Phi^\dag;\ldots]$.}
\eq
\label{action}
S=S_0+g^2\,S_{{\rm int}} \equiv S[\Phi;B,g,\theta]
\eqend
defined above obeys
\eq
\label{duality}
S[\Phi;B,g,\theta] = S\left[\tilde\Phi\,;\,B\,,\,\tilde g\,,
\tilde\theta\right] \ ,
\eqend
where
\eq
\label{scaling}
\tilde\Phi(x) =\sqrt{\Bigl|\det(B)\Bigr|}~\hat\Phi(B\,x)
\eqend
and
\eq
\hat\Phi(k) = (2\pi)^{-n}\,\int_{\reals^{2n}}d^{2n} x~\e^{-ik\cdot
x}\,\Phi(x) \ , \quad k\cdot x = k_\mu x^\mu
\eqend
is the Fourier transform of $\Phi(x)$. The transformed coupling parameters are
\eq
\tilde\theta =-B^{-1}\,\theta^{-1}\,B^{-1} \ , \quad
\tilde g =\Bigl|\det(B\theta)\Bigr|^{-1/2}\,g  \ .
\eqend
Moreover, the transformation $(\Phi;B,g,\theta)\mapsto(\tilde\Phi;B,\tilde g,
\tilde\theta)$ is a duality of the field theory, i.e.\ it generates a cyclic
group of order two.}
\bigskip

We will first establish the duality symmetry of $S_0$ with $m=0$. We use the
Parseval relation to rewrite $S_0$ in momentum space, i.e.\ in terms
of the Fourier transform $\hat\Phi$ of $\Phi$. We have
\eq
\widehat{(P_\mu \Phi)}(k) = \left(k_\mu- i\,B_{\mu\nu}\,
\hat\partial^\nu\right)\hat\Phi(k)
=\left(i\,\tilde\partial_\mu+B_{\mu\nu}\,\tilde k^\nu\right)\hat\Phi(k) \ ,
\eqend
where $\hat\partial^\mu = \partial/\partial k_\mu$, $\tilde k^\mu =
(B^{-1})^{\mu\nu}\,k_\nu $, and $\tilde \partial_\mu =
\partial/\partial \tilde k^\mu$. Thus the change of variables
$k\mapsto \tilde k$ yields
\beq
S_0 =\Bigl|\det(B)\Bigr|\,\int d^{2n}\tilde k~
\sqrt{\det (G) }~\left(G^{-1}\right)^{\mu\nu}\,
\left( \tilde P_\mu\hat \Phi
\right)^\dag\left(B\,\tilde k\right)\left(\tilde P_\nu\hat\Phi\right)
\left(B\,\tilde k\right) \ ,
\eeq
where
\beq
\tilde P_\mu = -i\,\tilde \partial_\mu-B_{\mu\nu}\,\tilde k^\nu \ .
\eea
Changing the name of the integration variable to $\tilde k=x$, this action has
same form as $S_0$ in \Ref{S0} except that $\Phi(x)$ is replaced
by the field $\tilde\Phi(x)$ defined in \Ref{scaling}. This proves that
$S_0[\Phi;B]=S_0[\tilde\Phi;B]$ for $m=0$. The proof of invariance
of the mass term
\eq
m^2\,\int d^{2n}x~\sqrt{\det (G) } \, \Phi^\dag(x) \Phi(x) \equiv
m^2\,\sqrt{\det (G) }~\langle \Phi ,\Phi\rangle
\eqend
is similar but simpler. Using the Parseval relation and performing the
transformation $k\mapsto x= B^{-1}\,k$ as above one finds $ \langle\Phi
,\Phi\rangle\, = \, \langle \tilde \Phi,\tilde \Phi\rangle $, as claimed.

We now turn to the proof the duality property of the action $S_{\rm
int}$. The crucial step is the behaviour of $S_{\rm int}$ under Fourier
transformation of the fields,
\eq
\label{crux}
S_{\rm int}[\Phi;G,\theta]=\Bigl|\det(\theta)\Bigr|^{-1}\,
S_{\rm int}\left[\hat \Phi\,;\,G\,,\,\theta^{-1} \right] \ ,
\eqend
i.e.\ Fourier transformation yields an action of the same form \Ref{Sint} but
with the Groenewold-Moyal product in momentum space defined as
\eq
\label{MPF}
\left(\hat f_1~\hat\star~\hat f_2\right)(k) =
\hat f_1(k)\,\exp\left(-i\,\overleftarrow{\hat \partial^\mu}
\,\left(\theta^{-1}\right)_{\mu\nu}\,
\overrightarrow{\hat \partial^\nu} \right)\,\hat f_2(k) \ .
\eqend
We first note that this readily implies the duality properties stated in
Section~1. Inserting $\hat\Phi(k)= |\det(B)|^{-1/2}\,\tilde\Phi(B^{-1}\,k)$ and
changing variables to $x=B^{-1}\,k$ gives
\eq
S_{\rm int}\left[\hat \Phi\,;\,G\,,\,\theta^{-1} \right] =
\Bigl|\det(B)\Bigr|^{-1}\,
S_{\rm int}\left[\tilde \Phi\,;\,G\,,\,-B^{-1}\,\theta^{-1}\,B^{-1}
\right] \ ,
\eqend
where we have inserted $\hat \partial^\mu= -(B^{-1})^{\mu\nu}\,\partial_\nu =
\partial_\nu\,(B^{-1})^{\nu\mu}$ into \Ref{MPF}. This yields
\eq
S_{\rm int}[\Phi;\theta] =
\Bigl|\det(\theta B)\Bigr|^{-1}\,
S_{\rm int}\left[\tilde\Phi\,;\,-B^{-1}\,\theta^{-1}\,B^{-1}\right] \ ,
\eqend
implying the third and fourth transformation rules in \Ref{11}.

We now prove \Ref{crux}. For this, we write $S_{\rm int}=
\sqrt{\det(G)}\,(\alpha\,S_\alpha+\beta\,S_\beta)$ and first treat the
action $S_\alpha$. This interaction term has a simple form in momentum
space,
\eq
S_\alpha=\prod_{a=1}^4\,\int\frac{d^{2n} k_a}{(2\pi)^n}~
\hat\Phi^\dag(k_1)\hat\Phi(k_2)
\hat\Phi^\dag(k_3)\hat\Phi(k_4)\,\hat \vv(k_1,k_2,k_3,k_4) \ ,
\eqend
with vertex function
\eqa
\label{vk}
\hat \vv(k_1,k_2,k_3,k_4)&=&\int d^{2n}  x~\left(\e^{-ik_1\cdot x}
\starx\e^{ik_2\cdot x}\right)
\left(\e^{-ik_3\cdot x}\starx\e^{ik_4\cdot x}\right)\nonu &=&
(2\pi)^{2n}\,\delta^{2n}(k_1-k_2+k_3-k_4)~
\e^{-i\theta^{\mu\nu}\bigl[(k_1)_\mu(k_2)_\nu +(k_3)_\mu(k_4)_\nu\bigr]} \ .
\label{vertex}
\eqaend
In a similar manner we can write the interaction term in position space,
\eq
S_\alpha=\prod_{a=1}^4\,\int\frac{d^{2n}x_a}{(2\pi)^n}~
\Phi^\dag(x_1)\Phi(x_2)\Phi^\dag(x_3)\Phi(x_4)\, \vv(x_1,x_2,x_3,x_4) \ ,
\eqend
and compute the position space vertex function by inverse Fourier
transformation,
\eq
\label{v4}
\vv(x_1,x_2,x_3,x_4)=\prod_{a=1}^4\,\int\frac{d^{2n}k_a}{(2\pi)^n}~
\e^{i(k_1\cdot x_1-k_2\cdot x_2+k_3\cdot x_3-k_4\cdot x_4)}
\,\hat \vv(k_1,k_2,k_3,k_4) \ .
\eqend
To appreciate the result of this integration, it is useful to recall that in
the conventional case ($\theta=0$) the interaction vertices in position
and momentum space are very different. The former one is fully local,
$\vv(x_1,x_2,x_3,x_4) \propto
\delta^{2n}(x_1-x_2+x_3-x_4)\,\delta^{2n}(x_1-x_2)\,\delta^{2n}(x_2-x_3)$,
while the latter one is non-local, $\hat \vv(k_1,k_2,k_3,k_4) \propto
\delta^{2n}(k_1-k_2+k_3-k_4)$.\footnote{\baselineskip=12pt A somewhat more
drastic example of this occurs in the chiral $p$-form field theories
studied in~\cite{pform}, in which duality maps a local model in
configuration space to a non-local one in momentum space.}
However, for non-singular $\theta$,
both vertex functions have the same non-local form. Using the representation
$(2\pi)^{2n}\,\delta^{2n}(k_1-k_2+k_3-k_4) = \int d^{2n}
t~\e^{-it\cdot(k_1-k_2+k_3-k_4)}$, the resulting momentum integrals in
(\ref{v4}) are all Gaussian, and a straightforward computation  yields (for
details see Appendix~A)
\eq
\label{vx}
\vv(x_1,x_2,x_3,x_4) =
\frac{(2\pi)^{2n}}{\Bigl|\det(\theta)\Bigr| } \,\delta^{2n}(x_1-x_2+x_3-x_4)~
\e^{-i(\theta^{-1})_{\mu\nu}\bigl[(x_1)^\mu(x_2)^\nu +(x_3)^\mu(x_4)^\nu
\bigr]} \ ,
\eqend
which, up to the factor $|\det(\theta)|^{-1}$, has the same form as the
momentum space vertex function \Ref{vk} but with the matrix $\theta$ replaced
by its inverse.  This implies that
$S_\alpha[\Phi;\theta]=|\det(\theta)|^{-1}\,S_\alpha[\hat\Phi;\theta^{-1}]$.

To see that the same property is true of the interaction term $S_\beta$, we
note that it can be written as
\eqa
S_\beta&=&\prod_{a=1}^4\,\int\frac{d^{2n}k_a}{(2\pi)^n}~
\hat\Phi^\dag(k_1)\hat\Phi^\dag(k_2)\hat\Phi(k_3)\hat\Phi(k_4)\,
\hat \vv(k_1,-k_2,-k_3,k_4)\nonu
&=&\prod_{a=1}^4\,\int\frac{d^{2n}x_a}{(2\pi)^n}~
\Phi^\dag(x_1)\Phi^\dag(x_2)\Phi(x_3)\Phi(x_4)\,
\vv(x_1,-x_2,-x_3,x_4) \ ,
\eqaend
with the {\it same} vertex functions $\vv$ and $\hat\vv$ as above. The
sign changes allowing us to use the results in \Ref{vk}, \Ref{v4} and
\Ref{vx} all work out since $k_1\cdot
x_1+k_2\cdot(-x_2)-k_3\cdot(-x_3)-k_4\cdot x_4 = k_1\cdot x_1
-k_2\cdot x_2 +k_3\cdot x_3-k_4\cdot x_4$.  This proves \Ref{crux}.
It is also evident that the transformation defined in Proposition~1 is
equal to its inverse, which thereby completes the proof of
Proposition~1.

\bigskip

\sect{3. Quantization} We will now show that this duality property holds true
at the full quantum level as well.  Formally, the quantum field theory is
defined by the functional integral
\eq
Z[J] = \int \cD\Phi~\cD\Phi^\dag~\e^{-S[\Phi;B,g,\theta]+
\langle\Phi,J\rangle + \langle J,\Phi\rangle} \ ,
\label{ZJdef}\eqend
where $\langle\Phi,J\rangle = \int d^{2n}x~\Phi^\dag(x)\,J(x)$ and
similarly for $\langle J ,\Phi \rangle$. Of course, a proper definition
requires a specification of ultraviolet and infrared regularizations, which
will be described below. Here the external source fields $J(x)$ and $J^\dag(x)$
are regarded as independent functions. The generating functional of all
connected Green's functions is then given by
\eq
{\cal G}(J)=-\ln\frac{Z[J]}{Z[0]}\equiv{\cal G}(J;B,g,\theta) \ .
\eqend
The path integral measure $\cD\Phi~\cD\Phi^\dag$ is defined so that for $g=0$, ${\cal G}(J)=\langle J,CJ\rangle$, where $C$ is the propagator of the quantum field theory which is given in position space by the free two-point correlation function
\eq
C(x,y) = \langle x|\left(P^2 + m^2\right)^{-1}|y \rangle \ .
\label{Cxy}\eqend

Since the functional integration measure is invariant under the transformation
$\Phi\mapsto \tilde\Phi$, and
\eq
\langle\Phi,J\rangle =
\int d^{2n}k~\hat \Phi^\dag(k)\,\hat J(k) =\Bigl|\det(B)\Bigr|^{-1}\,
\int d^{2n}k~\tilde \Phi^\dag\left(B^{-1}\,k\right)\,\tilde J\left(B^{-1}\,k
\right) =\left\langle \tilde \Phi\,,\,\tilde J\right\rangle
\eqend
along with the analogous property for $\langle J ,\Phi \rangle$, the change of
variables $\Phi\mapsto \tilde\Phi$ in the path integral and the duality
property of the action (Proposition~1) formally yield the identity
\eq
{\cal G}(J; B, g,\theta) = {\cal G}\left(\tilde J\,;\,B\,,\,\tilde g\,,
\,\tilde\theta\right) \ .
\eqend
This implies that any connected Green's function with $N$ external legs,
\eq
\cG_{n,N-n}(x_1,\ldots,x_N) = \left.
\prod_{c=1}^n \frac{\delta}{\delta J(x_c)}\,
\prod_{c=n+1}^N \frac{\delta}{\delta J^\dag(x_c)}
\, {\cal G}(J)\right|_{J=J^\dag=0} \ ,
\eqend
obeys the identity
\eq
\hat\cG_{n,N-n}(k_1,\ldots,k_N;B,g,\theta) =\Bigl|\det(B)\Bigr|^{N/2}\,
\cG_{n,N-n}\left(B^{-1}\,k_1\,,\,\ldots\,,\,B^{-1}\,k_N\,;\,B\,,\,\tilde g
\,,\,\tilde\theta\right)
\label{cGidentity}\eqend
where $\hat\cG_{n,N-n}$ is the Fourier transform of $\cG_{n,N-n}$,
\eq
\hat\cG_{n,N-n}(k_1,\ldots,k_N)
= \prod_{c=1}^n\,\int \frac{d^{2n}x_c}{(2\pi)^n}~\e^{-ik_c\cdot x_c}\,
\prod_{c=n+1}^N\,\int \frac{d^{2n}x_c}{(2\pi)^n}~\e^{ik_c\cdot x_c}\,
\cG_{n,N-n}(x_1,\ldots,x_N) \ .
\label{cGFourier}\eqend
The relation (\ref{cGidentity}) is particularly interesting for
$B\theta=\pm\,I$, since then $\tilde g=g$ and $\tilde\theta=-\theta$. In this
case,
Fourier transformation leaves all correlation functions invariant up to
rescaling and sign change of $\theta$.

To substantiate this argument we now show that there is a natural
regularization of the quantum field theory which cures all possible
divergences and which is also invariant under the duality transformation.
With this regularization included, all formal manipulations
above are put on solid ground, and they prove the stated conclusions
for the regulated Green's functions. The duality invariant regularization
amounts to replacing the propagator (\ref{Cxy}) by the following 
regulated one 
\eq
\label{ff}
C_\Lambda(x,y) = \langle x|\left(P^2 + m^2\right)^{-1}~\ff\left(\Lambda^{-2}
\,\left[P^2+Q^2\right]\right)|y \rangle \ ,
\eqend
where
\eq
\frac12\,\Bigl(P^2+Q^2\Bigr)\equiv\frac12\,\left(G^{-1}\right)^{\mu\nu}
\,\Bigl(P_\mu P_\nu +
Q_\mu Q_\nu\Bigr)=-\partial_\mu\,\partial^\mu + (Bx)_\mu\,(Bx)^\mu
\eqend
with $P_\mu=-i\,\partial_\mu-B_{\mu\nu}\,x^\nu$ (as in \Ref{P}) and
$Q_\mu=-i\,\partial_\mu + B_{\mu\nu}\,x^\nu$. The cut-off function $\ff(s)$ is
defined for real positive $s$ such that $\ff(0)=1$ (so that $C_\Lambda\to C$ as
$\Lambda\to\infty$) and such that it decays sufficiently fast as $s\to\infty$.
For definiteness, we will assume that $\ff(s)$ is smooth, monotonically
decreasing, and such that $\ff(s)=1$ for $0\leq s\leq 1$ and $\ff(s)=0$ for
$s\geq 2$. Heuristically, we would expect that this modification is enough to
regulate all potential ultraviolet and infrared divergences of the theory. The
term $-\partial_\mu\,\partial^\mu$ in the argument of $\ff$ cuts off the high
momentum modes, while the term $(Bx)_\mu\,(Bx)^\mu$ should regulate all
long-distance divergences. In Appendix~B we present the formal arguments which
substantiate this reasoning. Here we only show that this regularization is
indeed invariant under the duality transformation.

In Section~2 we showed that, under Fourier transformation, the differential
operator $P_\mu$ becomes $\hat P_\mu = k_\mu-i\,B_{\mu\nu}\,\hat\partial^\nu$,
and under the change of variables $k\mapsto x=B^{-1}\,k$ this transforms back
to
$-P_\mu$. Similarly, the duality transformation maps $Q_\mu \mapsto
-Q_\mu$. Thus
\eq
\ff\left(\Lambda^{-2}\,\left[P^2+Q^2\right]\right)~\longmapsto~\ff
\left(\Lambda^{-2}\,\left[P^2+Q^2\right]\right)
\eqend
under the duality for arbitrary cut-off functions $\ff$. This proves invariance
of the regularization procedure.

The results of this section can be summarized as follows.

\bigskip
\noindent {\bf Proposition~2 (Quantum duality):} {\it The
regularization defined above is invariant under the duality
transformation given in Proposition~1. Moreover, with this regularization and
for Euclidean spacetime metric $G=I$, all Feynman diagrams of the
quantum field theory defined by \Ref{action} can be
represented as {\em finite} sums and are thus convergent. The corresponding
regulated generating functional $\cG_\Lambda$ of all connected Green's
functions
is therefore well-defined, and it possesses the duality symmetry
\eq
\cG_\Lambda(J; B, g,\theta) =
\cG_\Lambda\left(\tilde J\,;\,B\,,\,\tilde g\,,\,\tilde\theta\right) \ ,
\eqend
where $\tilde J(x) =\sqrt{|\det(B)|}~\hat J(B\,x)$ and $\hat J$ is the
Fourier transform of $J$. }

\bigskip

\sect{4. Summary and discussion} We have considered a class of interacting
scalar quantum field theories in which the positions and momenta are
noncommuting coordinates. We showed that these models are invariant under
Fourier transformation, rescaling, and some specific changes of the coupling
parameters. These transformations generate the action of a group of order two
on the coupling parameter space and thereby define a duality of the quantum
field theory. This duality property was first shown on the level of classical
actions (Proposition~1) and then extended to the full quantum level
(Proposition~2). The quantum result was proven in perturbation theory, for the
special case of a trivial metric $G=I$, and only in the regulated quantum field
theory (with finite cut-offs). We conjecture that this result extends to the
cases with more general Euclidean metrics $G$ on $\real^{2n}$ and also beyond
perturbation theory. Along the way we have also made use of an interesting
technical aspect of these models, the fact that it is convenient to perform the
corresponding quantum field theoretic computations in `Landau space' (rather
than in position or momentum space), i.e.\ by expanding in the basis of Landau
wavefunctions $\phi_{\vn}$ as described in Appendix~B. This results in
somewhat unusual expressions for 
the Feynman diagrams. Similar techniques have also been applied
in~\cite{Habara,NCLandau}.

It would be interesting to study in detail what happens to this duality
property as the cut-off $\Lambda\to\infty$ is removed. The results presented
here can shed interesting light on the nature of divergences in noncommutative
quantum field theory, such as UV/IR mixing~\cite{UVIR}. In particular, the duality symmetry shows explicitly the well-known result that the correlation functions of the model do not have a continuous $\theta\to0$ limit. As discussed in
section~3, the regularization parameter $\Lambda$ can be regarded at the same
time as both an ultraviolet and an infrared cut-off. However, one can
generalize the regularization procedure by changing the argument of the cut-off
function $\ff$ in \Ref{ff} from $\Lambda^{-2}\,[ -\partial^2 + (B\,x)^2]$ to
$-\Lambda_{\rm UV}^{-2}\,\partial^2+\Lambda_{\rm IR}^{-2}\,(B\,x)^2$, i.e.\ use
different parameters $\Lambda_{\rm UV}$ and $\Lambda_{\rm IR}$ to cut-off high
momenta and long wavelengths, respectively. In conventional quantum field
theories one would interpret divergences which arise in the limit $\Lambda_{\rm
UV}\to\infty$ as ultraviolet divergences and those for $\Lambda_{\rm
IR}\to\infty$ as infrared divergences. However, in the present noncommutative
quantum field theory such a distinction between divergences appears to be
somewhat artificial. The duality transformation exchanges the parameters
$\Lambda_{\rm UV}$ and $\Lambda_{\rm IR}$ but otherwise leads to a model of
the same kind, i.e.\ ultraviolet divergences turn into infrared divergences
and vice versa. We expect that a more physical
interpretation of the divergences arises if one works consistently in position
space. We believe that this duality effect should also have some bearing
on the issue of renormalizability of this class of quantum field theories.

We conclude by pointing out a few further interesting aspects of these models:
\begin{itemize}

\item There is in fact a much larger group of symmetry transformations
associated with this class of noncommutative field theories. All
linear coordinate transformations $x\mapsto L\,x$, with $L$ an invertible
$2n\times2n$ real-valued matrix, map the field theory to one of the same kind
but with altered coupling parameters. In particular, the class of models with
$G=I$ are left invariant by orthogonal transformations $L\in O(2n,\real)$, but
of course both $\theta$ and $B$ change under these transformations.

\item For the class of models we have discussed, the special points 
$B\theta=\pm\,I$ are essentially but not quite self-dual since the
duality transformation changes the sign of $\theta$. There are, however,
simple variants of these quantum field theories which are {\it completely
self-dual}. Consider two flavours of scalar fields $\Phi_1$ and $\Phi_2$ and 
the total action $S'[\Phi_1,\Phi_2;B,g,\theta]=S[\Phi_1;B,g,\theta]+
S[\Phi_2;B,g,-\theta]$, with $S$ defined as in Section~2. Proposition~1
then implies that, if the duality transformation also involves an exchange $\Phi_1\leftrightarrow\Phi_2$ of the two scalar fields, the theory is fully
self-dual for $B\theta=\pm\,I$. These models in fact become more interesting
when one adds duality invariant terms which couple the two scalars, for
example a `mixed' mass term proportional to $\langle\Phi_1,\Phi_2\rangle+
\langle\Phi_1^\dag,\Phi_2^\dag\rangle$ or mixed interactions.

\item It is easy to see that Fourier transformation alone, without rescaling
the arguments of the fields, is also a duality transformation of the present
class of models. It acts on the parameters of the field theory as
\eqa
\label{FT}
B&\longmapsto& B^{-1} \ , \nonu
G&\longmapsto& -B^{-1}\,G\,B^{-1} \ , \nonu
\theta&\longmapsto& \theta^{-1} \ , \nonu
g&\longmapsto&\Bigl|\det(B\theta)\Bigr|^{-1/2}\,g\ ,
\eqaend
and now also affects the free part of the action. This duality can be regarded
as a generalization of the first equality in the harmonic oscillator identity
\Ref{1}. Note that in this transformation, the roles of covariant and
contravariant tensors on $\real^{2n}$ are interchanged, in contrast to the
previous duality transformation. This is a simple consequence of Fourier
transformation which maps $x^\mu\mapsto k_\mu$, and it is reminescent of what
occurs in standard T-duality transformations. In fact, at the special points
where $B=\pm\,\theta^{-1}$, the transformations \Ref{FT} are precisely those
obtained from the zero rank limit of the usual gauge Morita duality relations
between noncommutative Yang-Mills theories on $2n$-dimensional
tori~\cite{NCrev}--\cite{Morita}. This remarkable coincidence can be understood
by interpreting the noncommutative $\Phi_{2n}^4$-theory as a discrete $\zed_2$
noncommutative gauge theory~\cite{workinprogress}, and it is another sign of
the inherently stringy nature of noncommutative quantum field
theories~\cite{NCrev}.

\item In the special cases $\theta B=\pm\,I$ and $G=I$ the 
interaction vertices 
$\vvv(\vn_1,\vn_2,\vn_3,\vn_4)$ defined in Appendix~B simplify considerably,
and the resulting model has a natural interpretation as a matrix model. Based
on preliminary results~\cite{workinprogress} we conjecture that these models
are exactly solvable in $2n=2$ dimensions.
Moreover, the noncommutative soliton equations
derived from this model, including the free part of the action, at $\theta
B=\pm\,I$ are straightforward to solve, and the resulting soliton profiles are
the same as those which arise when the kinetic energy term is
neglected~\cite{GMS}.
\end{itemize}

\bigskip

\noindent
{\bf Acknowledgments}

\noindent
We thank C.-S.~Chu, J.~Gracia-Bond\'{\i}a, G.~Landi, U.~Lindstr\"om, F.~Lizzi,
M.~Salmhofer 
and K.~Zarembo for helpful discussions and correspondence. This work was
supported in part by the Swedish Science Research Council (VR) and the G\"oran
Gustafssons Foundation. The work of R.J.S. was supported in part by an Advanced
Fellowship from the Particle Physics and Astronomy Research Council~(U.K.).

\bigskip

\sect{Appendix A}
To compute the vertex function $\vv$ defined by \Ref{v4} and \Ref{vertex}, it
is convenient to introduce variables in $\R^{8n}$ defined by
\eq
K=(k_1,-k_2,k_3,-k_4) \ , \quad X=(x_1,x_2,x_3,x_4) \ .
\label{KXdef}\eqend
They allow us to write $\vv(X) = (2\pi)^{-4n}\,\int d^{8n}K~\hat
\vv(K)~\e^{-iK\cdot X}$, where
\eq
\hat \vv(K)= (2\pi)^{2n}\,\delta^{2n}(k_1-k_2+k_3-k_4)~
\e^{-\frac12\,K \cdot \A_\theta K }
\label{hatVK}\eqend
and $\A_\theta$ is the skew-symmetric $8n\times 8n$ matrix
\eq
\A_\theta = \left( \bma{rrrr} 0       & i\,\theta & 0 & 0 \\
-i\,\theta & 0      & 0 & 0 \\0&0&  0      & i\,\theta \\
0&0& -i\,\theta & 0 \ema\right) \ .
\eqend
Inserting $(2\pi)^{2n}\,\delta^{2n}(k_1-k_2+k_3-k_4) = \int d^{2n} t~
\e^{-iK\cdot T}$ with $T=(t,t,t,t)\in\R^{8n}$ and interchanging the order of
integrations gives the Gaussian integral
\eq
\vv(X)=\int d^{2n} t~\int\frac{d^{8n}K}{(2\pi)^{4n}}~
\e^{-i K \cdot (T+X) - \frac12\,K \cdot \A_\theta K }=
\det(\theta)^{-2}\,\int {d^{2n} t}~
\e^{ \frac12\,(T+X) \cdot \A_{\theta^{-1} }(T+X) } \ ,
\label{VXGauss}\eqend
where we have used $(\A_\theta)^{-1} = \A_{\theta^{-1} }$ and
$\det(\A_\theta) = \det(\theta)^4$. Now $T\cdot
\A_{\theta^{-1} }T = 0$ and $X\cdot \A_{\theta^{-1} }T = T\cdot
\A_{\theta^{-1} } X= -i\,(x_1-x_2+x_3-x_4)\cdot \theta^{-1}\,t$, and so the
$t$-integral in \Ref{VXGauss} yields
\eqa
\vv(X) = \det(\theta)^{-1}\,(2\pi)^{2n}\,
\delta^{2n}(x_1-x_2+x_3-x_4)~\e^{\frac12\,X \cdot \A_{\theta^{-1} } X} \ ,
\label{VXfinal}\eqaend
where one factor of $\det(\theta)^{-1}$ has been eliminated by the change of
variables $t\mapsto \theta^{-1}\,t$. From (\ref{hatVK}), and the fact that the
change of sign in the exponential in \Ref{VXfinal} is exactly compensated by
the different definitions of the vectors $K$ and $X$ in \Ref{KXdef}, it follows
that the expression \Ref{VXfinal} is identical to \Ref{vx}.

\bigskip

\sect{Appendix B} In this appendix we will prove that all Feynman diagrams
computed in the regulated quantum field theory, as defined in Section~3, are
given by finite sums. For simplicity we assume $G=I$. We believe that 
the extension of the argument to general constant Euclidean metrics $G$ is 
straightforward.

In the coordinate system where $B$ has the Jordan normal form
\eq
\label{J}
\left(B_{\mu\nu}\right)=\pmatrix{0&B_1& & & \cr-B_1&0& & & \cr
& &\ddots& & \cr & & &0&B_n\cr & & &-B_n&0\cr} \ , \quad B_j>0 
\eqend
we have
\eq
P^2\equiv P_\mu P^\mu = H_{12} + H_{34} + \ldots + H_{2n-1,2n} \ ,
\eqend
where, for each $j=1,\dots,n$, $H_{2j-1,2j}$ depends only on the spacetime
coordinates $x_{2j-1}$ and $x_{2j}$. It is identical to the Landau Hamiltonian,
i.e.\ the quantum mechanical Hamiltonian describing the motion of a particle in
the two-dimensional $(x_{2j-1},x_{2j})$-plane with an applied, constant
perpendicular magnetic field $B_j$. This Hamiltonian has well-known
eigenfunctions $\phi_{m_j,\ell_j}$ which are labelled by non-negative integers
$m_j$ and $\ell_j$, and obey $H_{2j-1,2j}\,\phi_{m_j,\ell_j} =
B_j(m_j+\frac12)\,\phi_{m_j,\ell_j}$. The eigenfunctions of the operator $P_\mu
P^\mu $ are therefore given by
\eq
\phi_{\vn}(x) = \prod_{j=1}^n\phi_{m_j,\ell_j}(x_{2j-1},x_{2j}) \ , \quad
\vn \equiv (\vm,\vl) = (m_1,\ldots,m_n,\ell_1,\ldots,\ell_n) \ ,
\eqend
with $P^2\,\phi_{\vn} = E(\vn)\,\phi_{\vn}$. One can check that they are also
eigenfunctions of the operator $Q^2$, $Q^2\,\phi_{\vn} = F(\vn)\,\phi_{\vn}$.
The eigenvalues are given by
\eq
E(\vn) = \sum_{j=1}^n B_j\left(m_j+\frac12\right) \ , \quad
F(\vn) = \sum_{j=1}^n B_j\left(\ell_j+\frac12\right) \ ,
\eqend
and they each depend on only half of the quantum numbers $\vn=(\vm,\vl)$. This
is just a generalization of the well-known degeneracy of the energy eigenstates
of the Landau problem. Note that complex conjugation of $\phi_{\vn}$ 
amounts to interchanging $\vm$ and $\vl$.

This basis allows one to diagonalize the free part of the action and is
therefore convenient for computing all Green's functions of the model, i.e. all
fields can be labelled by a $2n$-vector of non-negative integers $\vn$. By
expanding the scalar fields in this orthonormal basis,
\eq
\Phi(x) = \sum_{\vn} \phi^\dag_{\vn}(x)\,A(\vn) \ ,
\quad
\Phi^\dag(x) = \sum_{\vn} \phi_{\vn}(x)\,A^\dag(\vn) \ ,
\eqend
the free part of the action can be written as
\eq
S_0 = \sum_{\vn}\Bigl[ E(\vn) + m^2\Bigr]\,A^\dag(\vn)A(\vn) \ .
\eqend
The expansion of the interaction part of the action gives
\eq
\label{s0}
S_{\rm int}= \sum_{\vn_1,\vn_2,\vn_3,\vn_4}\vvv(\vn_1,\vn_2,\vn_3,\vn_4)\,
A^\dag(\vn_1)A(\vn_2)A^\dag(\vn_3)A(\vn_4) \ ,
\eqend
where the interaction vertices $\vvv(\vn_1,\vn_2,\vn_3,\vn_4)$ can be evaluated
by straightforward calculation from the position space vertex functions
$\vv(x_1,x_2,x_3,x_4)$ given in Section~2. The only property of these vertices
that we shall need for the present argument is that they are all well-defined.

{}From \Ref{s0} we see that the propagator in this representation is diagonal
and given by $c(\vn)= [ E(\vn) + m^2 ]^{-1}$. Moreover, the duality-symmetric
regularization amounts to replacing it by
\eq
c_\Lambda(\vn) = \frac{1}{E(\vn) +m^2 }~
\ff\left(\Lambda^{-2}\,\Bigl[E(\vn)+F(\vn)\Bigr]\right) \ .
\eqend
In this representation, all Feynman diagrams of the quantum field theory are of
the schematic form $\sum_{\vn'_1,\ldots,\vn'_K}\,\prod_kc_\Lambda(\vn'_k
)(\cdots)$, with $(\cdots)$ a product of interaction vertices $\vvv$, which
depend on the summation variables $\vn'_k$, and propagators $c$, which depend
on the labels $\vn_k$ of external legs that are not summed over. With the
assumed properties of the cut-off function $\ff$ stated in Section~3, the
propagator $c_\Lambda(\vn)$ is nonzero only if $E(\vn)+F(\vn) =
\sum_jB_j(m_j+\ell_j+1) < 2\Lambda^2$, which at finite $\Lambda$ is true for
only a finite number of distinct $\vn$'s. Thus all Feynman diagrams are
represented by finite sums. Note that these sums give the Green's functions
$\cG_{n,N-n}(\vn_1,\ldots,\vn_N)$ in the basis $\phi_{\vn}$. From
$\cG_{n,N-n}(\vn_1,\ldots,\vn_N)$ one can compute the corresponding position
space Green's function by multiplying it with the functions
$\phi_{\vn_k}(x_k)$, $k=1,\ldots,n$, and $\phi^\dag_{\vn_k}(x_k)$,
$k=n+1,\ldots,N$, and summing over all $\vn_k$. As above, one finds that these
sums are all finite and are thus well-defined, which completes the argument.

\end{document}